# A Multi Objective Reliable Location-Inventory Capacitated Disruption Facility Problem with Penalty Cost Solve with Efficient Meta Historic Algorithms

Elham Taghizadeh, Mostafa Abedzadeh, Mostafa Setak

*Abstract*— logistics network is expected that opened facilities work continuously for a long time horizon without any failure; but in real world problems, facilities may face disruptions. This paper studies a reliable joint inventory location problem to optimize cost of facility locations, customers' assignment, and inventory management decisions when facilities face failure risks and doesn't work. In our model we assume when a facility is out of work, its customers may be reassigned to other operational facilities otherwise they must endure high penalty costs associated with losing service. For defining the model closer to real world problems, the model is proposed based on p-median problem and the facilities are considered to have limited capacities. We define a new binary variable ($Z_{is}$) for showing that customers are not assigned to any facilities. Our problem involve a bi-objective model; the first one minimizes the sum of facility construction costs and expected inventory holding costs, the second one function that mention for the first one is minimizes maximum expected customer costs under normal and failure scenarios. For solving this model we use NSGAII and MOSS algorithms have been applied to find the pareto- archive solution. Also Response Surface Methodology (RSM) is applied for optimizing the NSGAII Algorithm Parameters. We compare performance of two algorithms with three metrics and the results show NSGAII is more suitable for our model.

*Keywords*—Joint inventory- location problem, facility location, NSGAII, MOSS

## I. Introduction

Recently most of the studies focus on facilities location problems, a large number of studies (e.g., Drezner. 1995; Owen and Daskin and Owen 1999) focused on the Uncapacited fixed charge location problem (UFL) that their goals is finding the optimal number of facilities and their locations in a supply chain network to balance the trade-off between facility setup costs and day to day shipment or transportation costs [1].However, in UFL problem inventory costs and the other were not usually considered. In many papers where product safekeeping is expensive, the holding cost and transportation cost may account for a significant portion of the total system cost. Utilizing UFL models to cases with significant inventory costs may yield suboptimal design and hug system cost estimation. Therefore researchers introduced joint inventory – location models that optimize facility locations to minimize the sum of the inventory costs, conclude the facility setup costs and the customer transportation costs. Various solution algorithms like lagrangian relaxation and column generation were used to solve the joint inventory-location models. Shu et all [2].

(2005) further improved these algorithms by exploiting certain special structures in the models. Meta –heuristics algorithms have also been used to solve these problems (e.g., Azad and Davoudpour, 2008) [1].

The facilities can be defined as fire station, emergency shelter, service center, logistics center and telecommunication post. The facility may provide service to one or several customer points. Traditional facility location often assumed that the facilities are always available and never incapacitate; they will provide service under any Conditions [3]. Many facilities are subject to potential operational disruptions from time to time. The famous ''lean'' concept is about allows development of global supply chains problems. From the terrorist attacks to the catastrophic devastation caused by Hurricane Katrina [4]. Many people believe that our international supply chains are strong and reliable. However in reality, these facilities can be unreliable; they will not provide service to customers that allocation to them because of maintenance, ranging from natural disasters to temporary shortages of capacity, breakdown or shut down for some unknown or known reasons. It is hence of theoretical and practical notification has been paid to facility location problems where facilities may not be completely reliable in recent years. When some facilities are not available, their customers either forced to travel excessive distances to access their demand or entirely give up the service and pay penalty [1]. Reliability is defined as the probability that a system or component performs its intended function within a given time horizon. A supply chain is reliable if it performs well when any parts of the system fail, for example, when a distribution center becomes unavailable due to some reason that we mentioned. Drezner presented the first mathematical models for facility location with unreliable suppliers, who studies the unreliable p-median and (p. q)-center location problems, in which a facility has a given probability of becoming failure. Snyder and Daskin provided two reliable facility location model formulations (based on p-median and UFL models) to research the effect of probabilistic facility failure on the optimal facility deployment [5]. They make the strong assumption that all facilities have the same probability of failure. Cui et al. developed their models to address site dependent facility failure probabilities in both discrete and continuous modeling [6]. Li and Ouyang further improved the continuum approximation model so as to solve problems under complex facility failure patterns [1]. Berman et al. (2007), Lim et al. (2009), Santivanez et al. (2009), Lim et al. (2010), Shen et al. (2007), Snyder et al. (2006) and Zhan et al. (2008) all consider models similar to Snyder and Daskin's but relax the uniform-disruption-probability assumption using a variety of modeling approaches. Oded Berman, Dmitry Krass introduce a new analytical approach that is based on representing the stochastic problem as a linear combination of deterministic p-median problems with facilities subject to

Elham Taghizadeh, Industrial Engineering Department, Khaje Nasir Tossi University, Iran, Tehran, Vanak sq., Molasadra Avenue, Number21(Elham_tgh@yahoo.com)

Mostafa Abedzadeh, Industrial Engineering Department, Khaje Nasir Tossi University, Iran, Tehran, Vanak sq., Molasadra Avenue, Number21(abedzade@kntu.ac.ir)

Mostafa Setak, Industrial Engineering Department, Khaje Nasir Tossi University, Iran, Tehran, Vanak sq., Molasadra Avenue, Number21(setak@kntu.ac.ir)

failure facing uniform demand. Snyder et al. provided reliable networks that perform as well as under normal conditions, while also performing well when disruptions. They are using p-robustness criterions for reducing the failure risks and solve their model with improved genetic algorithm [7].

Inventory control under supply chain disruption involves difficult nonlinear cost components, and such problems have been considered in recently (e.g., Ross et al., 2008; Qi et al., 2009; Schmitt et al., 2010). Some very recent studies, in the reliable location design framework, tried to develop models to address the joint inventory and facility location [1]. For example, Qi and Shen studied reliable delivery of finished products to satisfy stochastic customer demand when the supply chain is subject to random yield at the facilities [1]. Qi et al. further investigated the effects of facility failure at two supply chain echelons (e.g., one supplier and multiple retailers) on optimal retailer locations and customer allocations. Nevertheless, both studies assumed that a chain systems, if we allow customers to access backup services from other facilities (when their primary service facility has been disrupted) the supply chain system reliability and overall performance would be considerably improved. Qi Chen et al. proposed a nonlinear binary model to join inventory costs and a more general customer assignment scheme into the reliable facility location design framework. Their model optimizes facility location, customer allocations, and inventory management decisions when facilities are subject to disruption risks [1].

Recently most of studies try to use heuristicall and fuzzy algorithms to solve reliable and risk facility location problem for example, Nezir Aydin and Alper Murat (2012) use a hybrid method (a swarm intelligence based average approximation algorithm) for solving capacitated reliable facilities location problem (CRFLP)[3] and Babak H.Tabrizi and Jafar Razmi (2012) introduce a mixed-integer non- linear fuzzy model for solving risk management in designing supply chain networks Problem [3].

Most of Joint inventory – location models with reliable facility were not considered limited capacity for facilities. Hence, we propose in this paper a bi-objective Joint inventory location problem under the risk of probabilistic facility disruption with limited capacity for them same as in real world. Also we define another objective function that minimize maximum cost (time) of distance between facilities and customers. This objective function use for emergency facilities like fire station and so on. Briefly, our objectives in model are: 1. Cost of inventory and fixed charge located facilities, 2.Maximum cost of transportation based on distance between facilities and customers. The overall goal for this model is finding optimal solution by minimizing the both objectives. A number of case studies are conducted to test the proposed solution approach and draw insights on the optimal facility deployment.

The structure of this research is as follows: Section 2 introduces Assumptions and definitions, notation, the model formulation. Section 3 discusses about two methods of solving the problem. Section 4 conducts numerical experiments to test the proposed approach and show results. Finally in section 5, the Conclusion and Future Research are proposed.

## II. FORMULATION

### A. Assumptions and definitions

We consider the problem of optimally locating n facilities anywhere among a set of proposed location J; and allocating a set of customers I to them. Customers will assign nearby facilities for service and there aren't any limitations. For the unreliable new facilities, we should assume that the probability of a new facility becomes inactive is known. Each facility may fail with probability $q \in [0, 1]$. Facilities have independent failure probability and in the case where nearest facility failed; the next nearest facility will provide the service, etc. We assumed that when inventory of each facility is emptied, it orders.

We assume, when a facility fails it cannot provide any service and their customers will be either allocation to other active facilities or incur certain penalty (Qi Chen& etal, 2011). Customers can get service from a set of $R \leq |J|$ facilities. We assume in the normal Conditions (where no facilities fail) a customer is assigned to its level-0 facility. Whenever a customer's level –r facility fails (for any r≤R-1) it will be reassigned to its level-(r+1) facility (Qi Chen& etal, 2011). Indeed, facilities in the level-0 have backers in other levels. Note that due to independent failures, the probability for a customer to get service from its level-r facility is $(1-q)q^{r-1}$, i.e., the probability that its level-r facility is functioning while all lower – level facilities have failed (Qi Chen& etal, 2011). The probability for customer to incur penalty is $q^s$ the probability that all of its R assigned facilities have failed.

### B. Notations

Parameters
- I is the set of customers location
- J is the set of candidate location for facilities for servicing customers i∈ I.
- r index of level assignment
- $d_{ij}$ is the transportation cost from customer i∈ *I* to visit facility in location j∈ *J*.(also we can consider as distance between customer i and facility j)
- fj initial setup cost when a facility open in location j(installation and operation cost. fj> 0)
- q is probability that facility j is unreliable or out of service (fail)
- $p_{iu}$ is the penalty cost per unit of demand of customer i∈ I when all its R assigned facilities have failed.
- $\gamma_i$ is annual and given demand at customer location i∈ I.
- Kj maximum capacity of facility j∈ J
- |J| is the number of non-fail able facility
- hj is a holding cost per year for facility j∈ J (hj> 0).
- $b_j$ is a given order cost for a facility at j∈J (bj> 0).
- $P_j$ is a variable cost per unit of order ($p_j$> 0).

Decision variable
- The binary decision variables $X_j$, j∈ J determine the facility locations:

$X_j$=1, if a facility opens at location j; 0, otherwise.
- The binary decision variables decide how facilities are assigned to the customers:

$Y_{ijr}$ =1, if customer i is assigned to facility j at level r; 0, otherwise.

- The auxiliary binary variable :

$Z_{is}=1$, if customer i is assigned to unfaultable facility at level s, 0, otherwise.

*C. Formulation*

$$Min \sum_J \left[\left(2b_j h_j \sum_I \sum_{r=0}^{|J|-1} \gamma_i Y_{ijr}(1-q)q^r\right)^{1/2} + \sum_J \sum_{r=0}^{|J|-1} \gamma_i Y_{ijr} P_j (1-q) + f_j X_j\right]. \quad (1)$$

$$Min\, D. \quad (2)$$

s.t

$$D \geq \sum_{r=0}^{|J|-1} \sum_J \gamma_i Y_{ijr} d_{ij} q^r (1-q) + \sum_{s=0}^{r-1} \gamma_i p_{iu} q^s Z_{is}. \quad (3)$$

$$\sum_J Y_{ijr} + \sum_{s=0}^{r-1} Z_{is} = 1. \quad (4)$$

$$\sum_{r=0}^{|J|-1} Y_{ijr} \leq 1. \quad (5)$$

$$\sum_J \sum_{r=0}^{|J|-1} \gamma_i Y_{ijr} \leq K_j X_j. \quad (6)$$

$$Y_{ijr} \leq X_j. \quad (7)$$

$$X_j, Y_{ijr}, Z_{is} \in \{0,1\} \quad (8)$$

$$X_j, Y_{ijr}, Z_{is} \in \{0,1\}$$

The first objective function of this reliable joint inventory location problem (RJIL) minimizes the expected total cost (including the facility setup cost and the inventory cost) across the all possible facility failure scenarios. The second objective function is minimizing maximum expected weighted sum of distance or transportation cost from each customer to its nearest available facility. The transportation cost is assumed proportional to the moving distance between facilities. Almost the second objective is defined for problem with emergency facilities like ambulance, Fire station, Hospital and others. We generalize our model with defining second objective. In this model we want determining the optimal number of facilities and their locations with minimizing both objective functions. We want keep our model more reliable with minimum cost.

Compare of other models in reliability, in this model we define a new binary variable ($Z_{is}$) for showing the case that customers are not assigned to any facilities. Also this variable defines and considers penalty cost. For this propose we define level s that customers incur a penalty cost.

The first constraint show maximum transportation cost base on moving distance. Eq. (4) require that for each customer i at each level r is assigned to a facility j or it is assigned to a level-s facility (s < r) that is unfaultable. The Eq. (5) is forbidden a customer from being assigned to a given facility at more than one level. Eq. (6) allowed customers assigned to a facility until its capacity and it shows the limited capacity for facilities. Eq. (7) ensures that a customer can only assign to a location with an opened facility. The last Constraints define binary variable.

### III. METHODOLOGY

Our model is categorized as NP-hard problems, which cannot be solved by exact methods; hence we present a genetic algorithm and a scatter search algorithm for solving our proposed model and reaching to the objective which has been said before. In this step we explain steps of both algorithms in details.

*A. NSGAII*

The primary reason for choosing the NSGAII is their ability to find multiple Pareto-optimal solutions in one single run (Kalyanmoy Deb, 2001). Since the virtual reason why a problem has a multi-objective formulation is because it is not possible to have a single solution which simultaneously optimizes all objectives, an algorithm that gives a large number of alternative solutions lying on the Pareto-optimal front is of great practical value (Kalyanmoy Deb, 2001).

The Non-dominated Sorting Genetic Algorithm (NSGA) proposed in Srinivas and Deb was one of the first such evolutionary algorithms. Over the years, the main criticisms of the NSGA approach have been as follows:

High computational complexity of non-dominated sorting: The non-dominated sorting algorithm in use until now which in case of large population size is very expensive, especially since the population needs to be sorted in every generation (Kalyanmoy Deb, 2001).

We address all of these issues and propose a much improved version of NSGA which we call NSGA-II. From the simulation results on a number of hard test problems, we find that NSGA-II has a better advantage in its optimized solutions than PAES [8]. These results encourage the application of NSGA-II to more complex and real-world multi-objective optimization problems [8].

*B. Method*

The algorithm and steps of our NSGAII that used in this paper are as follow as:

Step1: (Initialization) Generate a set of random solutions for the initial population.

Step2: (Evaluation) calculates the two fitness functions values for each individual and finds nondominated solutions.

Step3 :( Density Estimation) calculates the average distance of two points on either side of this point along each of the objectives which was named crowding distance.

Step4: (survivor selection) Apply selection operation to the population together with newly generated individuals to build the next generation.

Step5: (crossover) Apply crossover operation to a prespecified percent of individuals selected from the population.

Step6: (Mutation) Apply mutation operation to a prespecified percent of individuals selected from the population.

Step7: (Termination) Repeat steps 2 to 6 steps until the termination criterion is met.

*C. Response Surface Methodology (RSM)*

The Response Surface Methodology (RSM) has been implemented for optimizing the genetic algorithm parameters. Let first introduce the RSM: RSM is a collection of statistical and mathematical techniques which is useful for developing, improving, and optimizing processes in which a response of interest is influenced by several variables and the objective is to optimize this response. RSM has important application in the design, development and formulation of new products, as well as in the improvement of existing product design. It defines the effect of the independent variables, alone or in combination, on the processes. In addition to analyzing the effects of the independent variables, this experimental methodology generates a mathematical model which describes

the chemical or biochemical processes. The RSM has been implemented in 3 following steps:
Step1. Preliminary determination of the independent parameters and their levels are carried out.
Step2. Selection the experimental design and prediction of the model equation.
 (a) Run proposed algorithm around center point and set a first-order model. If curvature test is significant, set a second-order model and go to step 3. Else go to (b).
 (b) Move sequentially along the direction of maximum decrease in the response when the response start to get worse stop and select last point as center, go to (a).
Step3. Determination of optimum point and check it.
In step 1, after a preliminary analysis of the NSGA algorithm, the three most commonly studied NSGA parameters, including population size, crossover rate and mutation rate are treated as design factors and the number of iterations that is for stopping condition is selected enough big level. After preliminary analysis the initial levels of parameters are selected, that are shown in table 1. The number of iterations is 60.

TABLE I
The Initial Level of GA Parameters

| Factor | Level |
| --- | --- |
| Population size | [30,50] |
| Crossover rate | [0.60,0.70] |
| Mutation rate | [0.15,0.25] |

In step 2 and by using Design Expert software, a 2-Level factorial design has been chosen and by selecting ranges of changing these 3 factors, 13 runs have been generated and for every run the genetic algorithm has been run 10 times and the mean function value has been calculated. Then a new range for new selected parameters has been selected and the steps of the 2-factorial design have been repeated again (13 runs, 10 NSGA runs for each), and then the software proposed the selected factors` values which are as follows:

TABLE II
Optimum levels of Parameters

| Factor | Level |
| --- | --- |
| Population size | 60 |
| Crossover rate | 0.7 |
| Mutation rate | 0.5 |

We have used these values for our factors in NSGAII algorithm.

### D. MOSS

SS is an example of the so-called evolutionary methods, with the difference (compared to other evolutionary methods) that its mechanism for searching is not based solely on randomization. SS is characterized by the use of a Reference Set (Ref_Set) of solutions. At each step reference solutions are combined to generate new solutions and update the current Reference Set according to some systematic rules.
The general steps of MOSS an algorithm is follow as:

Step1: (Initialization) Generate a set of random solutions for the initial population by genetic algorithm.
Step2: (Improvement Method) Apply one point crossover operation to a prespecified percent of individuals selected from the population for this step.
Step3 :( Reference Set Formation) for creating Ref-Set choose a set of random solution based on its size.
Step4: (Subset Generation Method) Create sub set generation method for subset generation reference set 1 and reference set 2.
Step5: (Solution Combination Method) Apply mutation operation to a prespecified percent of individuals selected.
Step6: (Reference Set Update Method) Replace improved solution and Update the Ref-Set.
Step7: (Termination) Repeat steps 2 to 6 steps until the termination criterion is met.
The brief Pseudo Code of MOSS algorithm that used in this paper is as follow as:
*Initialization:*
- Initialize N solutions as in it population by algorithm
- Initialize an empty set as Pareto archive.
- Out loop:
- Apply diversification method for current solutions.
- In loop:
  - Apply improvement method (crossover method)
  - Apply Pareto archive update method.
  - Apply reference set update method to construction reference set1 and reference set2
  - Apply sub set generation method for creation binary subset of reference set 1 and reference set 2.
  - Apply combination method (mutation method) on subset create by subset generation method.
- End IN loop
- End out loop

### IV. NUMERICAL EXPERIMENTS

In this section use two sets of numerical experiment for testing the proposed model and its solution approach. All the dataset are from Snyder and Daskin (2005): a 49 node set consisting of continental state capital cities as well as Washington, D.C.; an 88 node set consisting of the union of the 49 node set that mentioned and the set of 50 largest cities in the United States. Each city generates a customer demands are adjusted to the state population divided by $10^5$ for the 49 node set and to the city population divided $10^4$ for the 88 node set. The fixed cost $f_j$ is adjusted to the median home value in the city for both sets. The transportation cost $d_{ij}$ is equal to Squared Euclidean distance between i and j. the penalty costs $p_{iu}$ were drawn from uniform distribution: [1000. 7000] and $b_j$=1000. $P_j$=5. The holding cost $h_j$ equals $10^{-3}f_j$. The capacity of each facility $k_j$ were drawn from uniform distribution: [100. 1000].

*A. Comparison metrics*

For assessment quality and diversity of Metaheuristics algorithms, there are numerous and various comparison metrics. In this paper, we use two important metrics that are spacing(S) and diversification (D) metrics. They are following as:

$$S = \frac{\sum_{i=1}^{N-1}|\bar{d}-d_i|}{(N-1)\bar{d}}. \quad (9)$$

$d_i$ Is the Euclidean distance between any two consecutive solutions on the optimal boundary and $\bar{d}$ is average of these distances.

$$D = \sqrt{\sum_{i=1}^{n} \max(\|x_t^i - y_t^i\|)}. \quad (10)$$

The equation $\|x_t^i - y_t^i\|$ Represents the Euclidean distance between two adjacent solutions of $x_t^i$ and $y_t^i$ on the boundary is optimal.

*B. Results*

The model and solution approach are implemented in Matlab7.8 0 on a PC with 2.80 GHz CPU and 4 GB RAM. In the NSGAII algorithm, we set npop=60. Pm=0.5, pc=0.7, nc= ($\frac{pc \times npop}{2}$) ×2, Max-Iteration=10.The results for 49 node cases are summarized in table 2. In the MOSS algorithm, we set size of Ref-Set (B) =10, npop=6×B, size of Ref-Set1=0.7×B, Pm=0.5, pc=0.7, Max-Iteration=10. The results for 49 node cases are summarized in Table III.

In previous models the concept of reliability in facility location problems are considered the expected value of transportation costs. While in many cases, such as determining the location of emergency centers, distribution centers of perishable materials and fire station, aim to serve all customers in the minimum time. Minimize the expected value of distance of every customer location from emergency centers is not a perfect solution. Hence, minimizing the maximum distance of customers' location from the centers is considered. In our model we defined two objective functions; the first one is including the facility setup cost and the inventory cost and the second is maximum distance of customers' location from the distribution centers. We minimized both objectives in our model.

When serving customers took a long time, customers preferred not to give service from any facilities then the variables z is for these customers obtained one. So in our problems we incurred penalty cost for losing these customers. With the increasing the probability failure of facility, the number of customers who preferred not to give service and penalty cost increase proportionally

Also in Fig 1 for two example that solve with NSGAII algorithm, we show The total objective 1 increases continuously with q and problem sizes, because of the enormous additional cost incurred by customer reassignments. Based on table 3 and 5 we can show this result is repeated by MOSS algorithm.

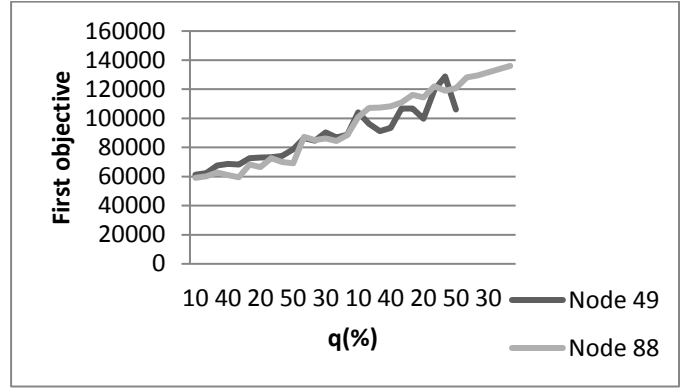

Fig .1. Relation between objective one and q& problem size (NSGAII algorithm)

Compared to Qi Chen et al (2011), in this paper we define another variable Zis for defining and considering penalty cost. We define level s that customers incur a penalty cost.

The numerical results for the 88 node datasets for two algorithms are displayed in Table IV&V. The results show the same behaviors like the 49 node dataset that presents in the example problem with 88 node dataset. Despite the increased problem sizes, the proposed solution approach can still solve our problem with acceptable results and solution time. As shown in tables 3 and 5, when compared with NSGAII algorithm, MOSS algorithm was able to find the same or not worse solution for 49 and 88 node datasets. For comparing these two algorithms, in our paper we calculated two comparison metrics that explained before.

We solved our model with two algorithms and show their results in our paper. In other section, for choosing best algorithm with maximum performance, we compare their results and analysis them.

*C. Compare NSGAII and MOSS*

We defined 11 problems and solved then with NSGAII and MOSS algorithms. As show in table 6 for obtaining clear analysis from our results we classified our problems in 11 groups.

For comparing Performance of two algorithm based on diversity metrics, we obtain value of this metrics (formulate 10) for two algorithms. Minimum metrics value for each algorithm, show its performance is better than the other algorithm. For spacing metric (formulate 9) maximum metrics value for each algorithm, show its performance is better than other algorithm.

For our model we solved each problem with defined parameters. For obtaining better solution solved each problem in several times (at least 10 repeated) and calculated two metrics for each times. As show in Table VI we obtained comparison metrics for each problem and each algorithm. For example in problems 1 and 2 based on diversity metrics in 60 percent of cases NSGAII algorithm obtain more better solutions than MOSS algorithm. The same analysis is repeated for other problems and summary of their results gathered in table VI. Finally based on our analysis the NSGAII algorithms obtained better solution for our model and it has best performance.

TABLE III
. Numerical results for 49 node dataset (NSGAII algorithm)

| N | q (%) | No. of pareto | solution time(s) | No.of facility | Obj-1 | Obj-2 | Diversity | spacing |
|---|---|---|---|---|---|---|---|---|
| 1 | 10 | 7 | 101.184972 | 5 | 61272.42 | 11920.77054 | 274.0570 | 0.6276383 |
| 2 | 20 | 5 | 96.092679 | 5 | 62286.98 | 1767.930677 | 189.9962 | 0.4308839 |
| 3 | 30 | 5 | 94.372176 | 5 | 67676.37 | 2228.631881 | 117.1565 | 0.446724 |
| 4 | 40 | 8 | 90.738473 | 5 | 68643.8 | 2377.078786 | 239.5423 | 0.876901 |
| 5 | 50 | 9 | 90.974415 | 5 | 68341.95 | 2517.305843 | 3453.9585 | 1.5594989 |
| 6 | 10 | 5 | 106.224825 | 7 | 72588.65 | 3196.050986 | 225.4825 | 0.7704151 |
| 7 | 20 | 7 | 100.706728 | 7 | 72986.43 | 5888.693956 | 3137.5280 | 1.639404 |
| 8 | 30 | 6 | 102.728175 | 7 | 73378.67 | 416050.25 | 2763.4538 | 1.6950158 |
| 9 | 40 | 12 | 98.951166 | 7 | 74247.81 | 11046.47 | 3994.1739 | 0.9948297 |
| 10 | 50 | 9 | 92.846686 | 7 | 78651.33 | 46648.83 | 3398.9173 | 1.4794568 |
| 11 | 10 | 8 | 112.245066 | 9 | 86339.29 | 4001.66 | 3281.9380 | 1.4147689 |
| 12 | 20 | 14 | 111.489371 | 9 | 84668.05 | 276953.89 | 4225.8299 | 1.3220533 |
| 13 | 30 | 8 | 113.881259 | 9 | 90330.45 | 4775.891679 | 3277.9708 | 1.4113515 |
| 14 | 40 | 13 | 110.204682 | 9 | 86748.34 | 547616.0057 | 3977.7141 | 1.576746 |
| 15 | 50 | 14 | 109.540698 | 9 | 88738.76 | 547906.7849 | 4038.6427 | 1.5298482 |
| 16 | 10 | 15 | 135.977374 | 11 | 104106.6 | 5370.940906 | 4358.3878 | 1.8652998 |
| 17 | 20 | 16 | 134.73926 | 11 | 96265.57 | 279481.3724 | 4365.8759 | 0.6558345 |
| 18 | 30 | 10 | 135.192585 | 11 | 91122.67 | 749514.4714 | 2757.6830 | 1.5113448 |
| 19 | 40 | 14 | 134.683594 | 11 | 93483.98 | 410027.0929 | 4051.3596 | 1.1780568 |
| 20 | 50 | 12 | 132.553583 | 11 | 106825.9 | 278275.5644 | 2854.3656 | 1.2969906 |
| 21 | 10 | 12 | 160.869818 | 13 | 106815.8 | 275810.4988 | 3723.7085 | 1.5299496 |
| 22 | 20 | 11 | 158.498465 | 13 | 99729.59 | 858261.8213 | 3731.2863 | 1.5377558 |
| 23 | 30 | 12 | 149.376874 | 13 | 119341.9 | 257142.3153 | 3718.1845 | 1.0543052 |
| 24 | 40 | 16 | 156.628816 | 13 | 80572.73 | 1023742.863 | 4406.0547 | 0.8687226 |
| 25 | 50 | 20 | 155.814126 | 13 | 106108 | 281078.0857 | 4875.7619 | 1.5316598 |

TABLE VI
Analysis of comparison metrics between two algorithms

| Problems | Solving Method | Diversity Metric | Spacing Metric |
|---|---|---|---|
| 1,2 | NSGAII | 60% | 50% |
|  | MOSS | 40% | 50% |
| 3,4 | NSGAII | 100% | 0% |
|  | MOSS | 0% | 100% |
| 5,6 | NSGAII | 80% | 20% |
|  | MOSS | 20% | 80% |
| 7,8 | NSGAII | 60% | 50% |
|  | MOSS | 40% | 50% |
| 9,10,11 | NSGAII | 60% | 80% |
|  | MOSS | 40% | 20% |

For comparing performance of two algorithms, another metrics can be use is the time that each algorithm took to obtain the optimum solution. For this propose in our paper we compared solution times for 49 nodes dataset in three problems (7, 9 and 11 facilities) and their results show in table VII and Fig 2.

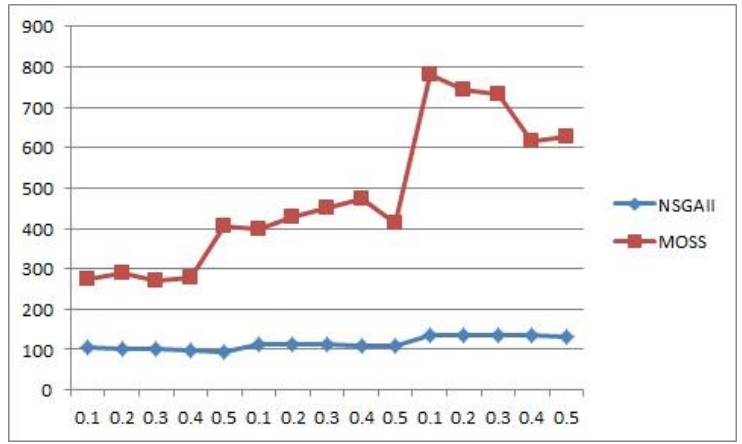

Fig 2: Compare Solution time(s) for 49 nodes dataset between two algorithms

Which algorithms with less solution time for same scale problems has better performance and solver prefers to choose this algorithm. As show in figure 6 their results show that NSGAII is more compatible and suitable for obtaining optimum solution of our model.

TABLE VII
Compare Solution time(s) for 49 nodes dataset between two algorithms

| 49 nodes | | | | | | | | |
|---|---|---|---|---|---|---|---|---|
| 7 Facilities | | | 9 Facilities | | | 11 Facilities | | |
| q | NSGAII | MOSS | q | NSGAII | MOSS | q | NSGAII | MOSS |
| 0.1 | 106.23 | 275.88 | 0.1 | 112.24 | 398.15 | 0.1 | 135.97 | 779.51 |
| 0.2 | 100.71 | 287.95 | 0.2 | 111.49 | 426.51 | 0.2 | 134.74 | 744.06 |
| 0.3 | 102.73 | 272.09 | 0.3 | 113.89 | 449.25 | 0.3 | 135.193 | 733.37 |
| 0.4 | 98.951 | 278.72 | 0.4 | 110.21 | 473.34 | 0.4 | 134.68 | 616.71 |
| 0.5 | 92.85 | 406.84 | 0.5 | 109.54 | 412.59 | 0.5 | 132.55 | 625.97 |

## V. CONCLUSION AND FUTURE RESEARCH

This paper proposes a reliable joint inventory facility location model that includes a general customer assignment mechanism and the inventory ordering and holding costs into the reliable facility location design framework with consider limited capacity for facilities. This model determines the optimal number of facilities, customer allocation and inventory management policies that minimize the expected inventory, customer and facility set up costs and minimize the maximums expected transportation cost across all possible facility disruption and in our model we introduce the new variable for missing our customers for the first time. We formulated a compact nonlinear program and developed a customized solution approach to efficiently obtain near optimum solutions. NSGAII and MOSS algorithms have been used for solving the proposed model. We used the Response Surface Methodology (RSM) for optimizing the NSGAII algorithm parameters. The computational results of applying the NSGAII and MOSS algorithms have been presented. Numerical results show that the proposed approach is able to obtain solutions in a short time under various problems settings. Managerial insights about the problem are drawn from these results. For example, we have found that customer demand tend to be joined together for service by only a few facilities when the inventory cost is dominating, while it will be spread to more facilities to reduce the shipment when the transportation cost is controlling over. When the facility failure probability increases, the expected total inventory and fixed setup costs and the number of constructed facilities both increase.

This work can be further extended in several directions. The use of lead time or backorders may affect supply chain structure and facility location design. In the real world, due to spatial heterogeneity and dependence of facility failure hazards, facility failure probabilities may present complex patterns such as site dependence and spatial correlation. It would be interesting to study how different facility failure patterns affect facility location design.

After these comparisons between two algorithms (NSGAII and MOSS) we concluded the NSGAII algorithm is more suitable and acceptable for our models with this dataset.

## APPENDIX

Appendix A: Table IV- Numerical results for 88 node dataset (NSGAII algorithm)
Appendix B: Table V- Numerical results for 88 node dataset (MOSS Algorithm)

TABLE IV
NUMERICAL RESULTS FOR 88 NODE DATASET(NSGAII ALGORITHM)

| N | q (%) | No.of pareto | solution time(s) | No.of facility | Obj-1 | Obj-2 | diversity | Spacing |
|---|---|---|---|---|---|---|---|---|
| 1 | 10 | 3 | 130.713657 | 5 | 59122.34 | 2248.96 | 100.56 | 0.0251373 |
| 2 | 20 | 4 | 134.380182 | 5 | 60242.97 | 4383.28 | 127 | 1.0811532 |
| 3 | 30 | 4 | 129.381309 | 5 | 62848.41 | 2866.34 | 196.67 | 0.4678745 |
| 4 | 40 | 6 | 126.433423 | 5 | 61041.23 | 5736.92 | 261.86 | 0.7108723 |
| 5 | 50 | 6 | 129.935916 | 5 | 59469.09 | 5735.94 | 234.10 | 0.6280393 |
| 6 | 10 | 4 | 161.099856 | 7 | 68154.05 | 3091.38 | 165.57 | 1.0192138 |
| 7 | 20 | 4 | 160.953274 | 7 | 66610.83 | 8918.58 | 191.16 | 0.2863371 |
| 8 | 30 | 3 | 164.606765 | 7 | 72601.01 | 2822.75 | 208.58 | 1.5342573 |
| 9 | 40 | 5 | 166.612876 | 7 | 70053.81 | 6437.18 | 206.21 | 0.3344347 |
| 10 | 50 | 5 | 159.255652 | 7 | 69240.8 | 5676.02 | 1047.31 | 1.8872246 |
| 11 | 10 | 3 | 210.598463 | 9 | 87389.37 | 2513.34 | 129.76 | 0.0555176 |
| 12 | 20 | 4 | 198.876889 | 9 | 85111.79 | 3140.99 | 161.01 | 0.2069076 |
| 13 | 30 | 4 | 195.296446 | 9 | 86173.33 | 3748.94 | 130.42 | 0.7295619 |
| 14 | 40 | 5 | 191.393809 | 9 | 84341.47 | 29202.91 | 561.35 | 1.0667824 |
| 15 | 50 | 5 | 192.820437 | 9 | 88515.81 | 3287.17 | 236.84 | 0.574728 |
| 16 | 10 | 7 | 241.676739 | 11 | 100898.3 | 4529.91 | 239.03 | 0.7638296 |
| 17 | 20 | 6 | 233.85527 | 11 | 107202.6 | 2862.57 | 236.17 | 0.4338384 |
| 18 | 30 | 6 | 236.78055 | 11 | 107439.4 | 3719.57 | 229.90 | 0.6351607 |
| 19 | 40 | 6 | 230.114748 | 11 | 108262.9 | 4201.83 | 182.80 | 0.569148 |
| 20 | 50 | 3 | 229.785389 | 11 | 110915.5 | 3093.86 | 152.60 | 0.331561 |
| 21 | 10 | 4 | 282.072033 | 13 | 116219.8 | 3221.16 | 175.79 | 0.1307587 |
| 22 | 20 | 5 | 278.559884 | 13 | 114405.2 | 3957.07 | 166.65 | 0.5605552 |
| 23 | 30 | 5 | 281.930554 | 13 | 122003.1 | 3338.35 | 240.82 | 0.6975716 |
| 24 | 40 | 5 | 282.465356 | 13 | 118921.4 | 7884.75 | 239.46 | 0.9610277 |
| 25 | 50 | 6 | 276.668113 | 13 | 120688.5 | 8077.07 | 175.19 | 0.9029292 |
| 26 | 10 | 5 | 350.234111 | 15 | 128152.3 | 5140.79 | 157.20 | 0.4378345 |
| 27 | 20 | 5 | 353.28214 | 15 | 129483.4 | 3757.15 | 174.28 | 0.4933476 |
| 28 | 30 | 5 | 338.075453 | 15 | 131679.2 | 4694.61 | 204.79 | 0.7511824 |
| 29 | 40 | 4 | 349.229145 | 15 | 133884.8 | 3783.43 | 145.48 | 0.7262069 |
| 30 | 50 | 5 | 344.557319 | 15 | 136050.2 | 6145.88 | 165.75 | 0.3997207 |

TABLE V

NUMERICAL RESULTS FOR 88 NODE DATASET (MOSS ALGORITHM)

| N | q (%) | No.of pareto | solution time(s) | No.of facility | Obj-1 | Obj-2 | diversity | Spacing |
|---|---|---|---|---|---|---|---|---|
| 1 | 10 | 6 | 402.695234 | 5 | 5154.588829 | 2809.304897 | 158.0479 | 0.4908 |
| 2 | 20 | 7 | 355.138019 | 5 | 4719.432706 | 3285.903412 | 152.0780 | 0.3833 |
| 3 | 30 | 6 | 372.633894 | 5 | 4567.393865 | 2890.46315 | 128.5606 | 0.9519 |
| 4 | 40 | 5 | 375.508847 | 5 | 4700.062453 | 2388.141582 | 135.5558 | 0.5285 |
| 5 | 50 | 9 | 333.74679 | 5 | 4198.031217 | 2715.090008 | 158.8588 | 0.2282 |
| 6 | 10 | 7 | 662.853849 | 7 | 4945.063288 | 2382.294507 | 171.0030 | 0.5247 |
| 7 | 20 | 7 | 14462.4891 | 7 | 4856.419663 | 3054.224243 | 149.6566 | 0.6223 |
| 8 | 30 | 6 | 682.60046 | 7 | 4997.179131 | 2753.816048 | 149.0160 | 0.6876 |
| 9 | 40 | 7 | 657.446953 | 7 | 4696.922304 | 2693.052924 | 155.4392 | 0.9757 |
| 10 | 50 | 6 | 629.609051 | 7 | 4483.07347 | 2600.149646 | 144.3037 | 0.9233 |
| 11 | 10 | 7 | 1182.44306 | 9 | 5038.566485 | 3000.973986 | 170.1766 | 0.5808 |
| 12 | 20 | 5 | 1284.57948 | 9 | 5485.426169 | 3508.825056 | 151.3494 | 0.7531 |
| 13 | 30 | 8 | 1213.29309 | 9 | 4949.000448 | 3571.127262 | 166.1868 | 0.9575 |
| 14 | 40 | 5 | 1090.26183 | 9 | 4764.032228 | 7299.959047 | 236.8037 | 0.3063 |
| 15 | 50 | 5 | 1144.39334 | 9 | 4725.519561 | 3242.675228 | 121.4829 | 0.6767 |
| 16 | 10 | 3 | 1985.93572 | 11 | 4949.885551 | 4127.536371 | 111.2970 | 0.0612 |
| 17 | 20 | 7 | 1751.47139 | 11 | 5135.736661 | 3226.356967 | 155.1206 | 0.521 |
| 18 | 30 | 7 | 1700.69361 | 11 | 5156.676073 | 3402.370086 | 159.8952 | 0.7098 |
| 19 | 40 | 12 | 1460.10846 | 11 | 5157.33905 | 4323.873386 | 225.6348 | 0.9849 |
| 20 | 50 | 6 | 1795.81849 | 11 | 5077.900956 | 5590.556012 | 232.7815 | 1.0769 |
| 21 | 10 | 9 | 2551.20904 | 13 | 4991.171919 | 3828.50444 | 271.4356 | 1.1349 |
| 22 | 20 | 7 | 2619.16897 | 13 | 5446.659333 | 4092.735877 | 217.4775 | 0.8683 |
| 23 | 30 | 5 | 7784.00058 | 13 | 5071.744472 | 3755.247513 | 132.0950 | 0.7995 |
| 24 | 40 | 5 | 2654.80647 | 13 | 5100.20206 | 3147.428161 | 130.1145 | 0.9301 |
| 25 | 50 | 9 | 2234.51107 | 13 | 5074.676025 | 4704.772357 | 294.0055 | 1.3152 |
| 26 | 10 | 5 | 3434.99546 | 15 | 5203.101167 | 3354.416997 | 124.9490 | 0.7337 |
| 27 | 20 | 7 | 2881.70127 | 15 | 5070.496077 | 4352.931219 | 177.8327 | 0.9638 |
| 28 | 30 | 5 | 3288.45577 | 15 | 5552.113742 | 3074.771922 | 157.0172 | 0.1521 |
| 29 | 40 | 4 | 3341.50782 | 15 | 5182.341047 | 3592.640377 | 114.9154 | 0.7931 |
| 30 | 50 | 7 | 5077.90258 | 15 | 5143.390982 | 3489.619006 | 155.1265 | 0.663 |